\documentclass[11pt,twocolumn]{article}
\usepackage{multirow}%

\usepackage{amsmath}
\usepackage{amsfonts}
\usepackage{graphicx}
\usepackage[superscript]{cite}
\usepackage{hyperref}
\usepackage{geometry}
\usepackage{titlesec}
\usepackage{fancyhdr}
\usepackage{xcolor}
\usepackage{subcaption}
\usepackage{svg}



\titleformat{\section}{\large\bfseries}{\thesection.}{1em}{}
\titleformat{\subsection}{\normalsize\bfseries}{\thesubsection.}{1em}{}

\title{\Large \textbf{The impact of spurious imaginary phonon modes on thermal properties of Metal-organic Frameworks}}

\author{
Prathami Divakar Kamath\textsuperscript{1,2}, Kristin A. Persson\textsuperscript{1,2,\color{blue}†}\\[0.5ex]
\textsuperscript{1}Department of Materials Science \& Engineering, University of California, Berkeley, CA, USA\\
\textsuperscript{2}Materials Sciences Division, Lawrence Berkeley National Laboratory, Berkeley, CA, USA\\
\textsuperscript{\color{blue}†} Corresponding author: kapersson@lbl.gov
}

\date{}

\begin{document}

\twocolumn[
\maketitle

\begin{center}
\large \textbf{Abstract}
\end{center}
\noindent
    Metal-organic Frameworks (MOFs) have emerged as potential candidates for direct air capture (DAC) of green house gases and water. Thermal properties of MOFs, such as their heat capacity, are used to determine the energy penalty associated with the adsorbent retrieval during the Temperature Swing Adsorption process. To aid exploration of the vast experimental design space of MOFs for such applications, computational methods like Density Functional Theory (DFT) or surrogate machine learning models trained on DFT data have been developed for obtaining phonon-derived heat capacities of MOFs. However, the high cost of explicit phonon computation in large and flexible nanoporous MOFs often necessitates the use of small supercells or lower convergence criteria which decrease predictive accuracy. These approximations often result in spurious imaginary phonon modes which are commonly ignored in practice. At present, there is no clear consensus in the literature on what magnitude of negative frequency or what fraction of imaginary modes can be considered acceptable. Here, we systematically demonstrate that spurious imaginary phonon modes can introduce substantial errors in heat capacity estimates, leading to incorrect ranking of MOFs in thermal-property-based screening. We further show that benchmarking machine learning interatomic potentials (MLIPs) against DFT datasets containing spurious imaginary modes can misrepresent models that predict physically meaningful phonon spectra for dynamically stable MOFs. Finally, we introduce a simple, rapid post-processing workflow that can be applied to standard phonon calculations to effectively correct heat capacity estimates and account for spurious imaginary modes in MOFs.

\textbf{Keywords:} Imaginary phonon modes, MOFs, Heat capacity, MLIPs, DFT
\vspace{1cm}
]

\section*{Introduction}

To enable the industrial deployment of Metal-organic frameworks (MOFs)\cite{mof,mof1,mof2,mof3}  in CO$_2$ adsorption-based nanotechnologies \cite{opendac,dac1,dac2}, heat-management costs during adsorbent regeneration through Temperature Swing Adsorption (TSA) processes \cite{tsa} must be considered alongside gas uptake performance \cite{Moosavi,hc1}. The total energy penalty associated with TSA-based regeneration comprises of two physically distinct contributions; The specific heat capacity C$_p$ of MOFs, which directly quantifies the sensible heat penalty, and the desorption enthalpy, related to the isosteric heat of adsorption $\Delta H_\mathrm{ads}$ \cite{luo2022experimental}. While both are important, the sensible heat contribution is particularly significant in the context of large-scale MOF screening for several reasons. Firstly, C$_p$ is a purely intrinsic framework property, independent of the choice of adsorbate unlike $\Delta H_\mathrm{ads}$, meaning that an accurate C$_p$ estimation is transferable across different gas separation applications without requiring recomputation. Secondly, in practical TSA processes where large volumes of absorbent material are cycled repeatedly, the total regeneration energy cost and efficiency was shown to be dominated by the sensible heat penalty of the adsorbent and therefore specific heat capacity represents one of the primary thermodynamic bottleneck for process efficiency \cite{kulkarni2012analysis,luo2022experimental}.

Accurately computationally predicting this property is therefore critical for realistic screening and design of promising MOFs. This task is complicated by the rich phonon landscape of MOFs\cite{softmodes}, originating from their hybrid organic-inorganic building blocks, flexible linkers, and large unit cells that host hundreds of vibrational modes. In particular, low-frequency soft phonons drive structural breathing and phase transitions \cite{breathing,Alhamami2014BreathingMOF,D3TA02214E,deform}, making phonon-quantized lattice vibrations \cite{phonons,Ziman2001} central to the thermodynamic stability and thermal properties of MOFs \cite{thermal}.

Computationally, phonons are typically calculated within the harmonic approximation \cite{harmonic} after a tightly converged structural relaxation, ensuring that atomic forces are effectively zero. In principle, imaginary phonon modes signal true dynamical instabilities, where small atomic perturbations grow exponentially and drive the system toward a lower-energy structure\cite{imag1}. In practice, however, spurious imaginary modes are ubiquitous in MOFs, arising from numerical artifacts such as insufficient structural relaxation, inadequate supercell sizes in finite-difference calculations, and broken translational symmetry \cite{imag}. The unusually high density of soft vibrational modes in MOFs \cite{softmodes,accidentalmodes} further amplifies this sensitivity, making the appearance of spurious, imaginary modes dependent on the chosen convergence criteria.

Obtaining high-fidelity phonon spectra for MOFs using density functional theory (DFT)~\cite{dft1,dft2,dft3} is therefore particularly challenging, as their large unit cells often require supercells containing thousands of atoms \cite{kamath,expphon,vmof}. Wieser \textit{et al.} \cite{Wieser2024} showed that near-elimination of imaginary modes is achievable for a small set of well-known MOFs by employing exceptionally tight convergence criteria and large supercells, but at a very high computational cost. While such settings yield highly reliable phonon spectra, they are impractical for large-scale studies involving hundreds of MOFs. Previous studies like \cite{imag,kamath,vmof} have proposed systematic strategies to eliminate spurious imaginary phonon modes, including mode mapping, atomic rattling followed by re-optimization, and tighter structural convergence. However, when applied to MOFs at the DFT level, these procedures can require significant additional computational hours per structure. As a result, despite the methodologies being well established in principle, they have seen limited practical adoption for MOF phonon calculations.

Consequently, most large-scale DFT phonon datasets for MOFs rely on looser convergence criteria and smaller simulation cells, where residual spurious imaginary modes are tolerated because their complete removal is computationally prohibitive. For example, recent efforts have produced valuable open-access DFT phonon datasets for tens to hundreds of MOFs, notably by Yue \textit{et al.} \cite{yue} and Moosavi \textit{et al.} \cite{Moosavi} where upto 6\% and 2\% imaginary modes were reported respectively. Notably, there is currently no clear consensus on what magnitude of a “small” negative phonon frequency can be considered negligible, with studies adopting different ad hoc cutoffs \cite{consensus1,consensus2,consensus3,consensus4,consensus5}. In practice,  all modes with imaginary frequencies are excluded from thermal property calculations, irrespective of whether their absolute magnitude is as small as 10$^{-4}$~THz or as large as $5$~THz. 

To extract maximal value from existing MOF phonon datasets, we argue that greater emphasis should be placed on the fraction of spurious imaginary modes that are omitted, rather than on the precise magnitude of their negative frequencies. We show that heat-capacity errors can substantially exceed the percentage of imaginary modes over certain temperature ranges, and that these two quantities cannot be directly equated. Moreover, neglecting imaginary modes can introduce non-negligible errors in heat capacity that lead to underestimated energy penalties and, consequently, to the mis-ranking of MOF candidates. We further caution against using DFT phonon data containing spurious imaginary modes as reference benchmarks for machine-learning interatomic potentials (MLIPs), as this practice can artificially penalize models that predict physically meaningful phonon spectra. Recognizing the prohibitive computational cost of eliminating spurious imaginary modes with fully converged DFT calculations, we propose a simple post-processing correction for heat-capacity estimation. This approach can be applied in seconds to standard phonon outputs from packages such as Phonopy \cite{TOGO20151}, yielding accuracy comparable to much more expensive calculations. Ultimately, we hope that our work provides guidance for assessing the impact of spurious imaginary modes on heat-capacity calculations and enables users to define application-specific error tolerances.
\section*{Results}
 \begin{figure*}[t]
    \begin{subfigure}[t]{0.45\textwidth}
    \centering
    \includegraphics[width=\textwidth]{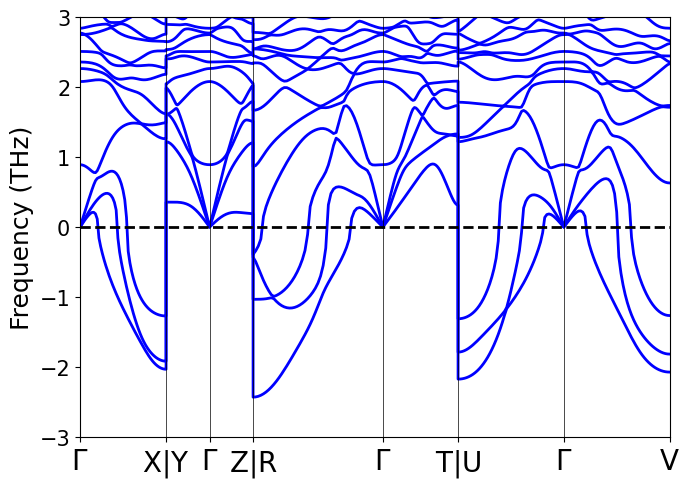}
    \caption{}
\end{subfigure}
\hfill
\begin{subfigure}[t]{0.5\textwidth}
    \centering
    \includegraphics[width=\textwidth]{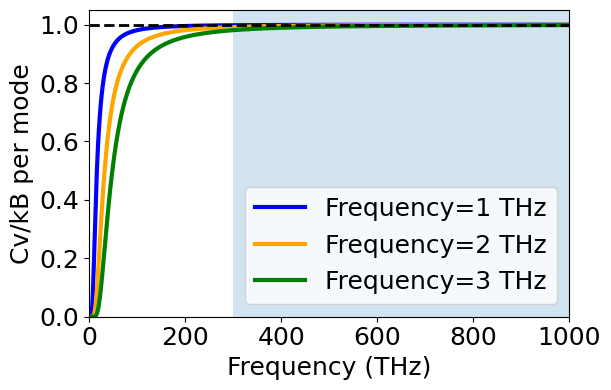}
    \caption{}
\end{subfigure}

\caption{(a) The phonon band diagram for MOF-74 obtained from the force constants reported in Moosavi \textit{et al.} \cite{Moosavi} using the primitive unit cell and relatively loose convergence criteria produces imaginary phonon modes for the branches corresponding to 3 acoustic and the lowest optical phonon mode. (b) The low-frequency modes (0-3 THz) which contribute significantly to thermal properties of MOFs can be approximated to be fully active at temperatures relevant for MOF screening applications. The area for temperatures $>$ 300K are highlighted as the region for interest for TSA applications}
    \label{fig:supp}
\end{figure*}

\subsection*{Effect of Spurious Imaginary Modes on Thermal Properties}
In this section, we illustrate the impact of a small fraction of spurious imaginary modes on the thermal properties of MOFs, focusing on two key quantities: the heat capacity and the contribution of low frequency phonon modes as a function of temperature.

Under the harmonic approximation, the heat capacity with Phonopy \cite{TOGO20151} is estimated with 
\begin{equation}\label{eq:one}
C_v(T) = k_B \sum_i \left( \frac{\hbar \omega_i}{k_B T} \right)^2 
\frac{e^{\hbar \omega_i / k_B T}}{\left(e^{\hbar \omega_i / k_B T} - 1\right)^2}
\end{equation} 

\begin{figure*}[t]
    \begin{subfigure}[t]{0.50\textwidth}
    \includegraphics[width=\textwidth]{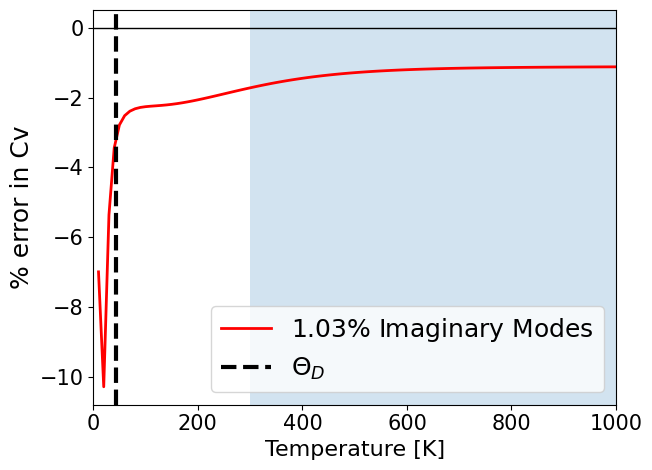}
     
     \caption*{\textbf{(a)}}
    \end{subfigure}
    \hfill
    \begin{subfigure}[t]{0.5\textwidth}
    \includegraphics[width=\textwidth]{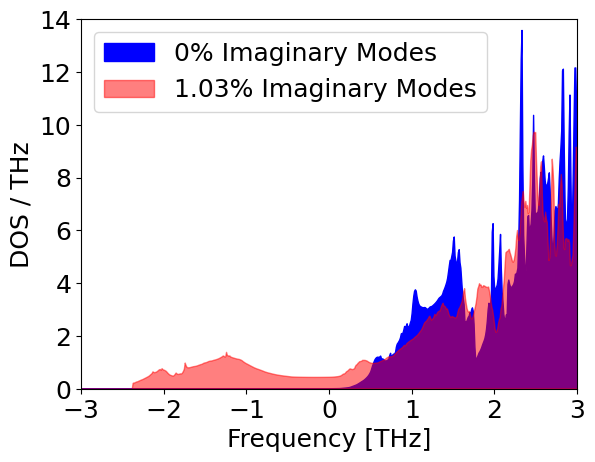}
       \caption*{\textbf{(b)}}
    \end{subfigure}
    \caption{(a)The \% deviations in $C_v$ obtained from DFT data with 1.03\% spurious imaginary modes relative to 0\% shows a strong temperature dependence. $\theta_D$ is the proxy Debye temperature of MOF-74 as per equation \ref{eq:debye} (b) The density of states (DOS) of the two phonon spectra obtained with and without spurious imaginary modes from Moosavi \textit{et al.} \cite{Moosavi} and Wieser \textit{et al.} \cite{Wieser2024} respectively on a $11\times11\times11$ mesh with the linear tetrahedron approach with a pitch of 0.01 THz.}
    \label{fig: cv}
\end{figure*}

Current practice for MOFs often assumes that imaginary frequencies can simply be omitted from Eq. \ref{eq:one}. However, a finite, temperature-dependent error in $C_v$ naturally follows from each neglected imaginary phonon mode. To determine whether this practice of omitting the imaginary modes is justified for MOFs, we first quantify this error in $C_v$ through a case study of a widely studied and dynamically stable MOF for gas adsorption applications, MOF-74 (with Zinc as the metal node) \cite{mof74}. The 1.03\% spurious imaginary modes observed in an underconverged phonon calculation for this MOF, as shown in Fig. \ref{fig:supp} a), corresponds to acoustic phonon branches predominantly and the lowest optical mode branch. Fig \ref{fig: cv} (a) demonstrates that spurious imaginary modes, even as small as 1.03\% can lead to an underestimation of $C_v$ by more than 10\% in the low temperature region and saturate to errors almost $2 \times$ the percentage of imaginary modes at $\sim$ 300K and higher temperatures.

The disproportionate errors in $C_v$ in Fig. \ref{fig: cv}(a) can be attributed to the nature of the imaginary modes. Fig. \ref{fig:supp}(b) directly demonstrates that phonon modes with frequencies up to $\sim$2 THz reach their full classical contribution of $k_\mathrm{B}$ per mode to $C_v$ well below 300K. For MOFs, this frequency region is dominated majorly by acoustic modes followed by the lower optical branches \cite{isophon,Wieser2024}, which are precisely the modes that are absent in the 1.03\% imaginary-mode curve of Fig. \ref{fig: cv}(b). Consequently, the percentage errors in $C_v$ are largest in the 0 to 50K range, where the contributions of the imaginary acoustic modes are omitted from the heat capacity calculation. The percentage errors in $C_v$ gradually decrease at higher temperatures as optical phonon modes increasingly contribute. Since the number of optical phonon branches scales as $3N-3$ (where $N$ is the number of atoms in the unit cell, typically $>50$ for MOFs), their cumulative contribution at higher temperatures rapidly exceeds that of the three acoustic phonons, reducing the relative impact of neglected imaginary modes on the calculated heat capacity.

To further supplement this discussion with a physical intuition for MOFs, we find that this low temperature range corresponding to largest errors in $C_v$ is also closely related to the Debye temperature. For complex systems such as MOFs with a large number of phonon branches, the estimation of the Debye temperature is non-trivial. Therefore, following the method in \cite{proxy} to determine a proxy for the true Debye temperature derived from acoustic modes, we use the frequency of the lowest optical mode at the $\Gamma$-point, having higher energy than all acoustic modes, to estimate ($\theta_D$) as defined in Eq.~\ref{eq:debye}. MOFs characteristically exhibit low Debye temperatures due to their soft and flexible nature \cite{huang2007thermal, mattesini2006ab} and the proxy estimates $\theta_D$ presented in this study in Table \ref{tab:debye} across different MOFs are solely intended to provide an intuition for this temperature scale. For MOF-74, this estimate yields $\theta_D \sim 43$ K. As per the Debye model \cite{mattesini2006ab,vykhodets2022debye}, for temperatures  $ T \gg \theta_D$, the acoustic modes vibrate in the classical temperature region where a full $k_B$ contribution to $C_v$ is achieved per mode. Regardless of whether selective capture of a targeted substance by MOFs is performed through ambient air DAC \cite{mcqueen2021review,shi2023temperature} or industrial point-source flue gas capture \cite{peh2022metal,zhao2019comprehensive}, the thermodynamically relevant temperature window for TSA-based adsorbent retrieval and heat capacity evaluation is broadly similar in both applications ($\sim$ 300–500 K). As shown in Fig. \ref{fig:supp} b), in this temperature range for TSA with MOFs, the majority of the low-frequency phonon modes, comprising of all the acoustic modes along with several optical modes, can be approximated to be fully active. Through the low frequency phonon contribution-based analysis in Fig. \ref{fig:supp} b) supplemented by the $\theta_D$ proxy discussion for physical intuition, the constant underestimation in $C_v$ in Fig. \ref{fig: cv} a) due to the neglected imaginary modes can now be quantified and rationalized for MOFs. 
\begin{figure*}[t]
    
    \begin{subfigure}[t]{0.50\textwidth}
         \includegraphics[width=\textwidth]{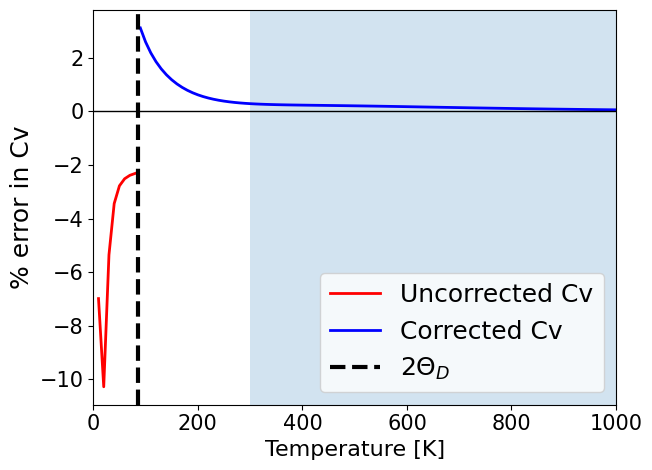}
         
        \caption*{\textbf{(a)}}
       
    \end{subfigure}
    \hfill
    \begin{subfigure}[t]{0.5\textwidth}
        
        \includegraphics[width=\textwidth]{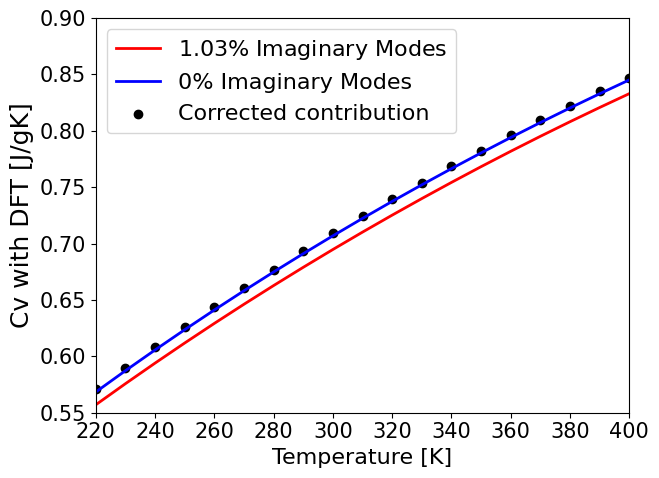}
        \caption*{\textbf{(b)}}
    \end{subfigure}
    \caption{(a)The \% deviations in $C_v$ obtained from DFT data with 1.03\% spurious imaginary modes relative to 0\% become significantly lower at $T \gg \theta_D$ after adding corrections (b) A constant underestimation $\sim 2\%$ can be seen in $C_v$ at temperatures around 300K due to the imaginary modes in the red curve. Accounting for the imaginary modes in the corrected contribution curve leads to a nearly perfect overlap with the blue 0\% Imaginary Modes Curve.}
    \label{fig: cv_cor}
\end{figure*}

Hence, we propose a simple remedy to account for the missing spurious imaginary modes by adding a correction term to the $C_v(T)_{\text{imaginary}}$ output from phonopy for high temperatures given by equation \ref{eq:corr}. 
\begin{equation}\label{eq:corr}
\begin{split}
C_v(T)_{\text{corrected}}
&= C_v(T)_{\text{imaginary}} \\
&\quad + k_B T \left( 3N \frac{\text{\% imaginary\_modes}}{100} \right)
\end{split}
\end{equation}

By assuming a full $k_b$ contribution at $T \sim 2\theta_D$ from the imaginary optical mode, this approach introduces a modest overestimation comparable in magnitude to the uncorrected data shown in Fig.~\ref{fig: cv_cor}(a). However, it also reduces the percentage error in $C_v$ from 1.7\% to 0.23\%, nearly an order of magnitude at 300 K and higher temperatures within our range of interest, resulting in an excellent agreement in Fig.~\ref{fig: cv_cor}(b). At temperatures around 1000 K where the high-frequency phonon spectra of the MOF can be expected to be activated, the overestimation in $C_v$ diminishes to zero. Thus, rather than simply omitting imaginary modes, this inexpensive post-processing correction can be readily applied and yields near-quantitative agreement with imaginary mode-free DFT results at moderate and high temperatures for heat capacity estimates.

 Owing to the high computational cost of high fidelity DFT calculations, we also report results obtained using MACE-MP-MOF0 \cite{kamath} in Table~\ref{tab:cv}. MACE-MP-MOF0 is a fine-tuned machine-learning interatomic potential (MLIP) designed for high-throughput phonon calculations of MOFs, for which tight convergence criteria can be readily applied to eliminate spurious imaginary modes. Given the excellent agreement between the specific heat capacity predictions from MACE-MP-MOF0 and DFT, this model is used as a surrogate for the 0\% imaginary-mode DFT baseline to extend the analysis to a broader and more diverse set of MOFs beyond MOF-74. 
\begin{table*}[]
    \centering
    \begin{tabular}{|c|c|c|c|c|}
    \hline
MOF  & Experimental C$_p$ & DFT $C_v$ & MACE-MP-MOF0 $C_v$ & MACE-MP-MOF0 C$_p$ \\ 
\hline
Zn-MOF-5 & 0.77 \cite{KLOUTSE20151}&0.79	&0.79&0.80	\\
\hline
Zr-UiO-66&0.78 \cite{10.1063/5.0201523}&0.78&0.78& 0.78\\
\hline
Zn-MOF-74&0.66 \cite{Moosavi}&0.71&0.71&0.71\\
\hline
Al-MIL-53&0.86 \cite{KLOUTSE20151}&1.01 &1.01&1.01\\
\hline
         
    \end{tabular}
    \caption{Constant pressure and volume specific heat capacity predictions (C$_p$ and $C_v$ respectively in $Jg^{-1}K^{-1}$) at room temperature. C$_p$ data with MACE-MP-MOF0 is obtained under the quasi-harmonic approximation. The DFT $C_v$ data was computed using the force constants from \cite{Wieser2024} as a reliable reference since there were $\sim$0\% imaginary modes. Refer the Methods Section for more details}
    \label{tab:cv}
\end{table*}
Table \ref{tab:cv} shows that, even when MLIPs reproduce DFT phonon properties with high fidelity, comparisons with experiment remain uncertain due to intrinsic DFT errors and the 1-3\% variability reported for the specific heat capacity of the same MOF across careful differential scanning calorimetry measurements \cite{RUDTSCH200217}. This challenge of experimental reproducibility is not unique to heat capacity measurements as similar variability has been reported in other aspects affecting gas adsorption efficiency such as low reproducibility of adsorption isotherms across laboratories for the same MOF \cite{park2017reproducible}, highlighting that experimental uncertainty is a pervasive challenge in the quantitative characterization of MOF performance. The specific heat capacity of MOFs is typically reported in the range of 0.6-1.4~$Jg^{-1}K^{-1}$ at room temperature \cite{hc}, consistent with the values obtained in our calculations (Table \ref{tab:cv}). Therefore, for two distinct MOFs that often differ by only $\sim$1-2\% in $C_v$, errors arising from spurious imaginary modes can match or exceed these uncertainties, thereby risking mis-ranking of MOF performance due to convergence criteria rather than fundamental limitations of DFT or MLIPs.

\subsection*{Systematic dependence of $C_v$ errors on \% Imaginary Modes}
\begin{figure*}[t]
\centering

\begin{subfigure}[t]{0.48\textwidth}
    \centering
    \includegraphics[width=\textwidth]{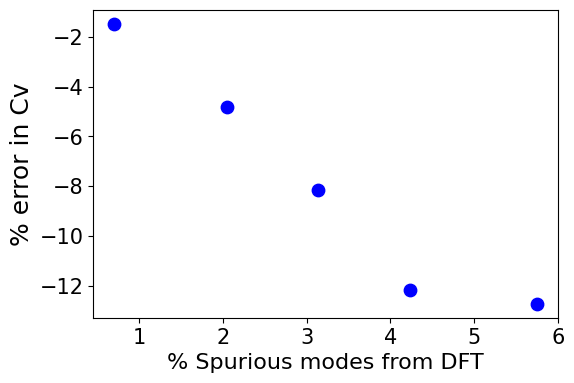}
    \caption{}
\end{subfigure}
\hfill
\begin{subfigure}[t]{0.48\textwidth}
    \centering
    \includegraphics[width=\textwidth]{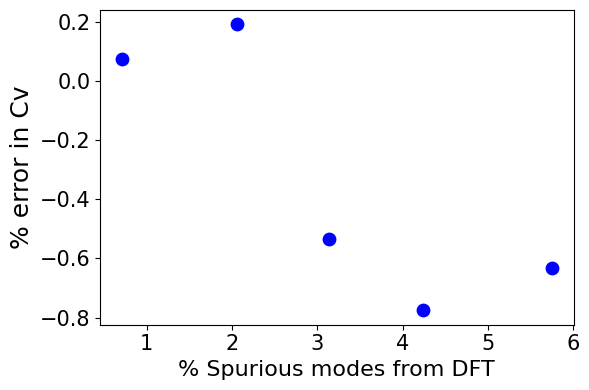}
    \caption{}
\end{subfigure}

\caption{(a) The deviations (\%) in $C_v$ obtained from DFT data with increasing spurious imaginary modes in different MOFs relative to the 0\% imaginary-modes data from MACE-MP-MOF0. 
(b) The deviations (\%) in the corrected $C_v$ obtained from post-processing the DFT data are $10\times$ smaller for all MOFs relative to the 0\% imaginary-modes data from MACE-MP-MOF0.}
\label{fig:cv_mof0}

\end{figure*}

In order to quantify the extent to which fraction of imaginary modes affect heat capacity predictions and to demonstrate the generalization of the proposed post-processing method across a diverse group of five MOFs, we investigate DFT phonon and $C_v$ data from Yue \textit{et al.} \cite{yue} (See Methods Section for details on the chosen MOFs). Using MACE-MP-MOF0 as a substitute for DFT for the 0\% imaginary mode baseline, Fig \ref{fig:cv_mof0} shows that DFT data with spurious imaginary modes varying from 0.71\% to 5.75\% leads to underestimations in $C_v$ at 300K from -1.5\% to -12.2\% respectively. This data indicates that the error in $C_v$ at 300K is more than 2 times the observed percentage of imaginary modes. Hence, for heat-capacity predictions of MOFs, maintaining spurious imaginary modes below a threshold of 0.5\% limits errors in  $C_v$
 to under 1\%, enabling reliable discrimination between distinct MOFs (as discussed in the previous section) without requiring post-processing corrections.

After adding the corrected contribution as per equation \ref{eq:corr} to the imaginary-mode DFT data, the percentage errors in $C_v$ decrease by an order of magnitude across all five examples demonstrating the robustness of the approach. Importantly, none of the selected MOFs were included in the training of the MACE-MP-MOF0 model. Therefore, the remaining errors after applying corrections to the DFT data may partially reflect the intrinsic accuracy of the model. Nevertheless, all deviations remain small, within an absolute error of 0.8\% 

\subsection*{Comparisons of DFT data with MLIP in the ranking of MOF Heat Capacity}
The rapidly expanding development of universal and fine-tuned MLIPs for MOFs has prompted increased efforts to benchmark their performance in predicting MOF physical properties. The analysis in the previous subsection suggests that even small spurious imaginary modes in DFT can artificially re-rank the heat capacities of MOFs relative to MLIP predictions with 0\% imaginary modes, potentially leading to the incorrect conclusion that an MLIP performs poorly. One recent example is MOFSimBench \cite{mofsim}, which compares the performance of 20 MLIPs in predicting MOF heat capacities relative to DFT reference data from Moosavi \textit{et al.} \cite{Moosavi}, which contain $\sim$2\% imaginary modes. The authors report a systematic overestimation of $C_v$ by MLIPs relative to DFT, attributing this behavior to softening of the potential energy surface in MLIPs~-~a phenomenon also observed in other studies \cite{soft}. 

\begin{figure*}[t]
    \centering
    \includegraphics[width=0.5\linewidth]{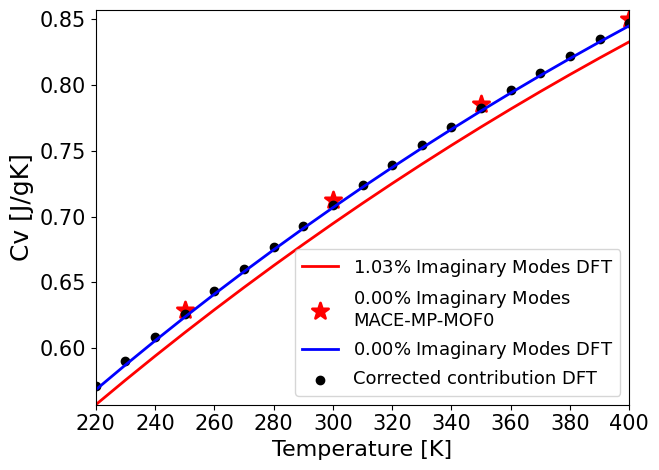}
    \caption{The $C_v$ curve from DFT data containing 1.03\% spurious imaginary modes shows a constant underestimation relative to other DFT \cite{Wieser2024} and MACE-MP-MOF0 data containing $\sim$ 0\% imaginary modes. The corrected contribution to the DFT data bridges this gap}
    \label{fig:mofsim}
    
\end{figure*}

We hypothesize that, in addition to this effect, a substantial contribution to the apparent MLIP overestimation arises from the presence of spurious imaginary modes in the reference DFT data, which are neglected in the heat capacity calculations. To test this hypothesis, we examine MOF-74 using MACE-MP-MOF0, which was identified as one of the top-performing models in MOFSimBench \cite{mofsim}, with a reported mean percentage absolute error (MPAE) of 2.5\% relative to DFT. As shown in Fig.~\ref{fig:mofsim}, MACE-MP-MOF0 overestimates $C_v$ by 2.53\% in agreement with the reported MPAE when benchmarked against the same DFT reference data containing 1.03\% imaginary modes used in MOFSimBench \cite{mofsim}. In contrast, when using DFT data with 0\% imaginary modes from Wieser \textit{et al.} \cite{Wieser2024}, as well as the corrected DFT data introduced in this work, the apparent overestimation by MACE-MP-MOF0 is substantially reduced to 0.77\%.
The MOFSimBench \cite{mofsim} study does not report whether imaginary modes were present and then neglected in the computation of $C_v$ across the 20 MLIPs considered, making it difficult to determine whether the observed overestimation reflect true model accuracy or compensating effects from imaginary modes in the MLIPs. (Refer SI Fig. S3 (c) for an example). 

\begin{figure*}[!htbp]
    
    \begin{subfigure}[t]{0.50\textwidth}
         \includegraphics[width=\textwidth]{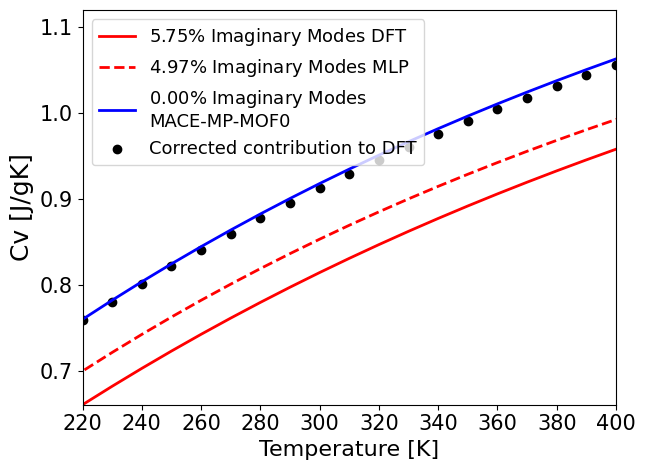}
         
        \caption*{\textbf{(a)}}
       
    \end{subfigure}
    \hfill
    \begin{subfigure}[t]{0.5\textwidth}
        
        \includegraphics[width=\textwidth]{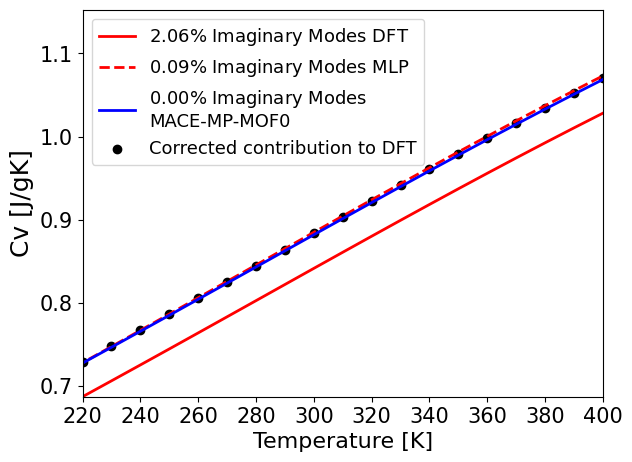}
        \caption*{\textbf{(b)}}
    \end{subfigure}
    \caption{The $C_v$ curves obtained from DFT with and without corrections for imaginary modes compared against MACE-MP-MOF0 and the MLP from Yue \textit{et al.} \cite{yue} for (a) "qmof-08b5558" and (b)"qmof-574737f"}
    \label{fig: cv_mlp}
\end{figure*}

While benchmarking MLIP performance or identifying best-performing models is not the primary focus of this work, we present an illustrative example demonstrating how models containing spurious imaginary modes can appear to exhibit lower errors when benchmarked against insufficiently converged DFT reference data. To this end, we compare MACE-MP-MOF0 with another recently fine-tuned MLIP for phonon and heat capacity prediction in MOFs proposed by Yue \textit{et al.} \cite{yue} using the reported DFT reference data in that study. We analyze two Zn-based MOFs from the previous section drawn from the QMOF database \cite{qmof1,qmof2}, which exhibit 5.75\% (ID ``qmof-08b5558'') and 2.06\% (ID ``qmof-574737f'') spurious imaginary modes in the DFT phonon spectra. For the MOF ``qmof-08b5558'' shown in Fig. \ref{fig: cv_mlp}(a), the Yue \textit{et al.} \cite{yue} (referred to as \textbf{MLP} in both their work and ours), \textbf{MLP} predicts 4.97\% imaginary modes, whereas MACE-MP-MOF0 yields zero imaginary modes. If the reference DFT data are used without correcting for spurious imaginary modes in the $C_v$ calculation, MACE-MP-MOF0 would be characterized as the inferior model, exhibiting an apparent deviation of $\sim$12.7\% at 300~K. In contrast, when the post-processed DFT $C_v$ data is used, MACE-MP-MOF0 achieves near-quantitative agreement with a deviation of 0.5\%, while the Yue \textit{et al.} \cite{yue} \textbf{MLP} shows a $\sim$10$\times$ larger deviation of 6.5\% relative to the corrected DFT reference. For the MOF ``qmof-574737f'' shown in Fig. \ref{fig: cv_mlp}(b), where the Yue \textit{et al.} \cite{yue} \textbf{MLP} also predicts $\sim$0\% imaginary modes, both MLIPs show improved agreement with the post-processed DFT $C_v$ data, with errors of $\sim$0.18\% at 300~K. These results underscore the importance of accounting for spurious imaginary modes in benchmarking analyses, as avoidable, model- and reference-data-specific errors can otherwise bias screening and ranking outcomes.
 \subsection*{Regime of Validity and Limitations}
 Flexible and breathing MOFs such as those of the MIL-53 family \cite{breathing} represent a subclass of particular relevance to the correction scheme proposed here. Their characteristically soft, low-frequency phonon modes and flexible linker networks make them disproportionately susceptible to spurious imaginary modes at the DFT level. As Wieser et al. \cite{Wieser2024} demonstrated that fully converged phonon spectra for the narrow-pore phase of MIL-53 could not be achieved even with extremely tight convergence criteria and large supercells due to prohibitive computational cost, making an inexpensive post-processing correction especially valuable for this technologically important subclass. We note, however, that for breathing MOFs, care must be taken to distinguish numerically spurious imaginary modes from those reflecting genuine dynamical instabilities or large-amplitude anharmonic motion associated with structural phase transitions for which the harmonic correction proposed here is not appropriate and more advanced treatments would be required \cite{imag}.
 
The limitations of the proposed correction scheme and the broader analysis presented in this work should be acknowledged. The post-processing correction \ref{eq:corr} is derived under the classical equipartition approximation and is therefore valid only in the temperature regime where $\hbar\omega \ll k_\mathrm{B}T$. While Fig. \ref{fig:supp} b) demonstrates that this condition is well-satisfied for the low-frequency modes observed in the datasets analyzed here at temperatures relevant to TSA-based gas capture applications (300–500K), the correction should not be applied uncritically at cryogenic temperatures where a full k$_B$ contribution can lead to significant overestimation. Second, the correction scheme does not distinguish between spurious imaginary modes arising from numerical underconvergence and imaginary modes that reflect genuine dynamical instabilities. Finally, while the correction restores the the missing vibrational contributions to heat capacity estimates, it does not address other sources of spectral inaccuracy in underconverged phonon calculations such as missing phonon density of states in the real frequency region or the long-range interactions that affect the acoustic branch dispersion in band diagrams.
    
We emphasize that this work does not endorse the presence of imaginary modes as acceptable practice, nor does it suggest that underconverged phonon calculations are sufficient for properties beyond heat capacity. Many thermodynamic and mechanical properties remain sensitive to the full quality of the phonon spectrum in ways that cannot be similarly corrected: vibrational entropy and free energy are acutely sensitive to the entire low-frequency spectrum due to the logarithmic frequency dependence of the harmonic oscillator entropy \cite{grimme}, rendering even small errors in soft-mode frequencies thermodynamically significant \cite{kundu2016ab} as also shown in SI Fig. 4 ; and elastic constants and mechanical stability criteria require accurate acoustic branch dispersions that are among the most severely affected by supercell underconvergence \cite{imag}. The proposed scheme is therefore offered as a practical resort to salvage heat capacity estimates from existing underconverged datasets when full recomputation is prohibitive. We encourage these datasets to be extended to other properties only with careful case-by-case justification. Looking forward, we regard the development and deployment of MLIPs capable of producing fully ab-initio quality converged, physically meaningful phonon spectra as the most promising path toward reliable, high-throughput phonon calculations in MOFs.

\section*{Discussion}
This work systematically elucidates the impact of spurious imaginary phonon modes arising from common computational approximations, such as insufficient structural relaxation, inadequate supercell sizes in finite-difference calculations, and broken translational symmetry~—~issues that are particularly prevalent in MOFs due to their large unit cells and computationally demanding relaxations. Although such imaginary modes are generally regarded as artifacts, their contributions are often neglected without a systematic assessment of their influence on calculated thermal properties. 

Rather than prescribing a single universally “correct” threshold for acceptable imaginary modes in phonon calculations, we present an analysis that quantifies the consequences of specific methodological choices on the physical properties of MOFs, enabling users to assess the impact on their intended applications and make informed decisions about how to treat observed imaginary modes.  With accurate, imaginary mode-free phonons~-~or with strategies to eliminate spurious imaginary modes that are more readily achievable using modern MLIPs due to their computational efficiency~-~we argue that MLIP benchmarking should go beyond property-level agreement and explicitly verify the physical validity of the underlying phonon spectra. This distinction between spurious imaginary modes and genuine dynamical instabilities is essential for obtaining reliable derived properties and for overcoming the scalability limitations of DFT. When fully imaginary mode-free DFT phonon data for MOFs are not feasible, the proposed post-processing workflow provides an efficient and practical alternative for benchmarking MLIPs against DFT.  The accompanying code is publicly available on GitHub and requires only standard Phonopy output files as input, enabling improved accuracy with negligible additional computational effort. 

Ultimately, this work demonstrates how approximations at the nanoscale vibrational level propagate to macroscopic thermophysical properties. Neglecting spurious imaginary phonon modes leads to systematic underestimation of heat capacity, which in turn results in a comparable underprediction of the energy required for MOF regeneration in CO$_2$ temperature swing adsorption processes. Such errors can distort materials screening and ranking, causing some candidates to appear more efficient than they truly are.
\begin{table*}[!htbp]
    \centering
    \begin{tabular}{|c|c|c|c|c|c|}
    \hline
    QMOF ID & No. of atoms in&No. of atoms in& MACE-MP-MOF0& $\theta_D$ & \%  DFT imaginary \\
    & DFT unit cell & MACE-MP-MOF0 cell &Supercell matrix  & (K)& modes \cite{yue}\\
    \hline
    qmof-f75eb5a &34 &612 &[3,0,0],[0,2,0],[2,2,3]& 72.76&0.71 \\
    \hline
    qmof-574737f &38 &1026  &[3,0,0],[0,3,0],[1,1,3] & 37.57& 2.06 \\
    \hline
    qmof-b12c68c &40 & 1440 & [4,0,0],[0,3,0],[1,0,3]& 88.97&  3.13\\
    \hline
    qmof-fab2f1a &43 & 1161&[3,0,0],[2,3,0],[1,2,3] & 61.79& 4.24\\
    \hline
    qmof-08b5558 &39 & 1404 & [3,0,0],[2,4,0],[0,0,3] &84.73& 5.75\\
    \hline
    \end{tabular}
    \caption{The table summarizes the key characteristics of the five MOFs from the QMOF database \cite{qmof1,qmof2} that were studied in Yue \textit{et al.} \cite{yue} to produce DFT harmonic phonon data using the primitive cell as the supercell which results in spurious imaginary modes. Also provided are the large supercells containing thousands of atoms used in the MACE-MP-MOF0 harmonic phonon calculations~—~which yielded 0.00\% imaginary modes~—~highlighting the scale of supercells required to accurately capture MOF lattice dynamics, a regime that is computationally prohibitive for conventional DFT calculations. Here, $\theta_D$ is the Debye temperature proxy computed with MACE-MP-MOF0. We refer SI Fig. S1 for the structural details of these MOFs}
    \label{tab:debye}
\end{table*}

\section*{Methods}
\subsection*{Phonon and Heat Capacity Computations}
The DFT force constants used to compute heat capacities in this work which were taken from Moosavi \textit{et al.} \cite{Moosavi}, Yue \textit{et al.} \cite{yue}, and Wieser \textit{et al.} \cite{Wieser2024} were all computed using the PBE+ D3(BJ) \cite{pbegga,d3,bj} functional and the trained MLIPs, MACE-MP-MOF0 \cite{kamath} and the \textbf{MLP} from Yue \textit{et al.} \cite{yue} were also trained on datasets generated with the same DFT functional ensuring consistency in our comparisons. The v2 version of the MACE-MP-MOF0 model was used in this work. The constant volume heat capacity $C_v$ data reported in this study was computed under the harmonic approximation with the Phonopy \cite{TOGO20151} package, using the force constants on a $11 \times 11 \times 11$ mesh for temperatures from 0 to 1000K. The \% spurious modes are computed from the number of modes neglected in the thermal properties calculation with cutoff frequency set to zero. 

 Among the thermal properties included in the standard output of Phonopy which are the vibrational Helmoltz free energy (F), Entropy (S) and Energy (U) in addition to Heat Capacity ($C_v$), $C_v$ has been the main focus here since it is a more directly measurable property reported in experimental literature for MOFs. This allows us to correlate the impact of computational artifacts like spurious imaginary modes to real-world applications such as energy penalty estimations from heat capacity. But to understand how the error in $C_v$ from spurious imaginary modes propagates through other derived thermodynamic variables, refer the SI Fig. S4. In order to compare experimental C$_p$, the quasi-harmonic approximation \cite{harmonic} was used to introduce the volume dependence indirectly with harmonic computations at $\pm$ 3\% linear strain for seven configurations around the equilibrium volume. Because of the expensive nature of this approximation with DFT for MOFs, we generated C$_p$ data only with MACE-MP-MOF0. The negligible difference seen between C$_p$ and $C_v$ in Table \ref{tab:cv} allows for more large scale analysis of the heat capacity of MOFs as the significantly cheaper harmonic phonon approximation can be used. 

The harmonic phonon computations with MACE-MP-MOF0 were performed by starting with a tight structural relaxation until the maximum force on the atoms is $10^{-8}$ \text{eV/\AA} and creating a supercell with minimum \text{20\AA} length to capture long-range interactions for the finite difference approach. The density of states and thermal properties were computed with the aforementioned mesh. None of the resultant harmonic phonon spectra from MACE-MP-MOF0 for the MOFs considered in this study needed elimination of any spurious imaginary modes. The compressed structures generated from the equilibrium configurations for the QHA calculations resulted in some imaginary modes which were eliminated (\% imaginary modes $\leq$ 0.08\%) using the atom-rattling and molecular dynamics procedure proposed in \cite{kamath}. Refer SI Section 4 for further discussion of the observed imaginary modes in the QHA calculations.

\subsection*{Chosen MLIPs and MOFs for the benchmarking example}
The MACE-MP-MOF0 and Yue \textit{et al.} \textbf{MLP} \cite{yue} were chosen to demonstrate an example of benchmarking as both models included training datasets spanning several equilibrium and out-of-equilibrium configurations of MOFs and have been previously studied for their phonon performance. While MACE-MP-MOF0 covers different metal-node based MOFs, since the other model only covers zinc (Zn) metal nodes, the chosen MOFs are Zn-based with diverse linkers and topologies to be compatible with both models as summarized in Table \ref{tab:debye}. Furthermore, to consider a wide range of \% imaginary modes for systematic evaluation, phonon spectra for all 40 MOFs sampled by Yue \textit{et al.} \cite{yue} from the QMOF database \cite{qmof1,qmof2} were analyzed to select five unique \% values as shown in Table \ref{tab:debye}. 

\subsection*{Estimation of the Debye Temperature}
The phonon spectra of MOFs are characterized by several low frequency acoustic and optical phonon modes due to their soft and flexible nature as shown in the phonon band diagrams of MOF-74 in Fig. \ref{fig:supp} a) and the five MOFs considered in this study (refer SI Fig. S2). The lowest optical mode frequency at the $\Gamma$ point ($\omega_\text{opt}^\text{min}$) serves a quick proxy and upper limit to the Debye cutoff frequency ($\omega_D$) of the acoustic phonons as determined in \cite{proxy}. This provides a rough estimation of the Debye temperature ($\theta_D$) for framework structure according to the equation
\begin{equation}\label{eq:debye}
    \theta_D = \frac{\hbar \omega_D}{k_b} \sim \frac{\hbar \omega(\Gamma)_\text{opt}^\text{min}}{k_b}
\end{equation}

Table \ref{tab:debye} summarizes the $\theta_D$ estimates from MACE-MP-MOF0 with $\sim$ 0\% imaginary modes for several MOFs to provide an idea about the range of Debye temperatures of MOFs and that the corrected contribution can be safely applied at 300K and higher temperatures. In cases where the phonon band diagrams contain spurious imaginary modes, $\theta_D$ should not be estimated directly as the acoustic phonon branches would be shifted below the 0 THz frequency limit which can lead to nonphysical predictions of temperatures. This proxy is not needed as an input for using the correction scheme and is used rather than a rigorous estimation of the Debye temperature, as the purpose of this computation is to show that the room temperature and higher ranges are far greater than the range of $\theta_D$ for MOFs.
\section*{Data Availability Statement}
The code for the proposed solution as well as the DFT and MLIP outputs have been made available on Github \url{https://github.com/prathami11/Spurious_ImaginaryModes/tree/main}
\section*{Acknowledgments}
This research used resources of the National Energy Research Scientific Computing Center (NERSC), a Department of Energy Office of Science User Facility. PDK acknowledges financial support from U.S. National Science Foundation's "The Quantum Sensing Challenges for Transformational Advances in Quantum Systems (QuSeC-TAQS)" program.
\section*{Author Contribution} All authors contributed towards the conceptualization of the work. PDK led the investigation, generated all the data for the results reported and drafted the manuscript. KAP supervised the research and guided data interpretation and assisted with drafting the manuscript.
\section*{Supplementary Information}
Supplementary data is uploaded as an ancillary file accompanying this article.

\end{document}